\documentclass[showpacs,aps,prd,nofootinbib,floatfix,amsmath,amssymb]{revtex4-1}
\usepackage{graphicx}
\usepackage{multirow}
\usepackage{xcolor}

\begin{document}

\title{Production of charged Higgs $H^\pm$ through   $cb$-fusion at LHC}

\author{J. Hern\'andez-S\'anchez}
\email{jaime.hernandez@correo.buap.mx}
\affiliation{Fac. de Cs. de la Electr\'onica, Benem\'erita Universidad Aut\'onoma de Puebla, Apartado
 Postal 1152, 72570 Puebla, Puebla, M\'exico}
 \affiliation{Dual CP Institute of High Energy Physics, C.P. 28045, Colima, M\'exico}
 \author{C. G. Honorato }
\email{carlosg.honorato@correo.buap.mx}
\affiliation{Fac. de Cs. de la Electr\'onica, Benem\'erita Universidad Aut\'onoma de Puebla, Apartado
 Postal 1152, 72570 Puebla, Puebla, M\'exico}
 \author{S. Moretti}
\email{s.moretti@soton.ac.uk}
\email{stefano.moretti@physics.uu.se}
\affiliation{School of Physics and Astronomy, University of Southampton, Highfield, Southampton SO17
 1BJ, United Kingdom,and Particle Physics Department, Rutherford Appleton Laboratory, Chilton, Didcot,
 Oxon OX11 0QX, United Kingdoma}
 \affiliation{Department of Physics and Astronomy, Uppsala University,
Box 516, SE-751 20 Uppsala, Sweden}
\author{S. Rosado-Navarro}
\affiliation{Fac. de Cs. F\'{\i}sico-Matem\'aticas, Benem\'erita Universidad Aut\'onoma de Puebla, Apartado Postal 1364, C.P.  72570 Puebla, Puebla, M\'exico} 
\email{sebastian.rosado@protonmail.com}
\begin{abstract}
We consider the framework of the 2-Higgs Doublet Model Type III (2HDM-III), wherein two doublets are coupled  to both up and down fermions and Flavour Changing Neutral Currents (FCNCs) are controlled by a four-zero  texture approach in the Yukawa matrices. We study the production of charged Higgs bosons ($H^\pm$) by means  of $cb$-quark fusion followed by the decay channel  $H^\pm \to \tau^\pm \nu$. Taking into account all experimental bounds as well as theoretical constraints, we show that, in a lepton-specific-like incarnation of the model, we obtain a significant sensitivity to such a process at the Large Hadron Collider (LHC). We come to this conclusion after a thorough Monte Carlo (MC) analysis comparing the aforemetnioned signal to  both  irreducible and reducible backgrounds from the SM.  

\end{abstract}

\maketitle

\section{Introduction}

The Standard Model (SM) limit of Electro-Weak Symmetry Breaking (EWSB) dynamics induced by a Higgs potential exists in several Beyond the SM (BSM) extensions of the Higgs sector. The 2HDM \cite{Branco:2011iw}
in its  Types I, II, III (or Y)   and IV (or X), wherein FCNCs mediated by (pseudo)scalar Higgs states can be eliminated under discrete symmetries \cite{Branco:2011iw}, is one notable example. However, another, equally  interesting  2HDM is  the one where FCFNs can be controlled by a particular texture in the Yukawa matrices \cite{Fritzsch:2002ga}. This model has a phenomenology that is very rich (see, e.g., Refs.~\cite{DiazCruz:2009ek}--\cite{HernandezSanchez:2013xj}),  like flavour-violating quarks decays. 
Furthermore, in this BSM scenario, the parameter space can avoid many of the current experimental constraints from flavour  and Higgs physics, so that a light charged Higgs  boson  (i.e., with a mass below  the top quark one) is allowed \cite{HernandezSanchez:2012eg} and  the  decay $H^- \to b \bar{c}$  can have a dominant Branching Ratio (BR), much larger than those of the  flavour diagonal $s\bar c$ and $\tau\nu$ channels. (Notice that this channel  has been  studied in a variety of Multi-Higgs Doublet Models (MHDMs) \cite{Akeroyd:2016ymd,Akeroyd:2012yg}, wherein the BR$ (H^- \to b \bar{c}) \approx 0.7- 0.8$ and one could obtain a considerable gain in sensitivity to the $H^\pm$ presence by tagging the $b$ (anti)quark).  An experimental framework to search for this kind of signal has been investigated in \cite{Slabospitsky:2002gw}.

In this work, by exploiting such an enhancement of the $H^- \to c\bar{b}$ vertex and  building upon the results previously presented in Ref.~\cite{Hernandez-Sanchez:2020vax}, we study the production of a light charged Higgs state at the  LHC  via heavy-quark fusion $b \bar{c} \to H^- $ followed by the decay  $H^- \to \tau \bar \nu_\tau$ (hereafter, charge conjugated channels are always  implied), with the $\tau$ decaying leptonically. We investigate this process in the framework of the aforementioned 2HDM-III with so-called lepton-specific couplings  and assess the LHC sensitivity to this  production and decay dynamics  against the irreducible background $q \bar{q}'  \to W^- \to \tau \bar \nu_\tau$ as well as the reducible noise due to $g {q}'  \to W^{\pm} q$ (with an additional jet) and $q \bar{q}  \to W^{+} W^{-} \to l^{+}l^{-}\nu \nu$ (where one lepton escapes detection). 
\section{The 2HDM-III}

As explained in \cite{Hernandez-Sanchez:2020vax} (and references therein), owing to the presence of a four-zero Yukawa texture as the mechanism to control FCNCs, the Yukawa sector of the 2HDM-III does not need a  discrete symmetry  \cite{DiazCruz:2009ek}--\cite{HernandezSanchez:2013xj}. Therefore, the Higgs potential for both doublets is given in the most general form,
\begin{eqnarray}
V(\Phi_1, \Phi_2) & = & \mu_1^2 (\Phi_1^\dag \Phi_1) +  \mu_2^2 (\Phi_2^\dag \Phi_2) -  \mu_{12}^2 [(\Phi_1^\dag \Phi_2) + h.c.] \nonumber \\
& + & \frac{1}{2} \lambda_1 (\Phi_1^\dag \Phi_1)^2+ \frac{1}{2} \lambda_2 (\Phi_2^\dag \Phi_2)^2+  \lambda_3 (\Phi_1^\dag \Phi_1)(\Phi_2^\dag \Phi_2)+ \lambda_4 (\Phi_1^\dag \Phi_2)(\Phi_2^\dag \Phi_1) \nonumber \\
&+ &\bigg(\frac{1}{2} \lambda_5 (\Phi_1^\dag \Phi_2)^2+  \lambda_6 (\Phi_1^\dag \Phi_1)(\Phi_1^\dag \Phi_2)+ \lambda_7 (\Phi_2^\dag \Phi_2)(\Phi_1^\dag \Phi_2) + h.c. \bigg),
\end{eqnarray}
where the doublets $\Phi_i=( \phi^-, \phi_i^{0*})$ ($i=1,2$) have hypercharge $+1$ and all parameters of the Higgs potential are real, including the Vacuum Expectation Values (VEVs). In such a scenario, the Yukawa sector is given by
\begin{eqnarray}
\mathcal{L}_Y = \bigg(  Y_1^u \bar{Q}_L \tilde{\Phi}_1 u_R + Y_2^u \bar{Q}_L \tilde{\Phi}_2 u_R+  
Y_1^d \bar{Q}_L \Phi_1 d_R + Y_2^u \bar{Q}_L \Phi_2 d_R + Y_1^\ell \bar{L}_L \tilde{\Phi}_1 \ell_R + Y_2^\ell \bar{Q}_L \tilde{\Phi}_2 \ell_R\bigg),
\end{eqnarray}
where $ \tilde{\Phi}_i= i \sigma_2 \Phi_i^*$  ($i=1,2$) and the two doublets are coupled with both up and down type fermions. After EWSB and the diagonalisation of the fermion matrices we can obtain, as a good approximation,  the rotated matrices $\bar{Y}_n^f $ by means of
\begin{eqnarray}
 \bar{Y}_n^f  = \frac{\sqrt{m_i^f  m_j^f} }{v} \chi_{ij}^f,
\end{eqnarray}
being the $\chi's$ parameters constrained by flavour physics, discussed widely in  \cite{HernandezSanchez:2013xj}. So, the interactions of charged Higgs bosons with fermions are given by
\begin{eqnarray}
\mathcal{L}^{\bar{f}_u f_d H^+}= -\bigg( \frac{\sqrt{2}}{v} \bar{u}_i (m_{d_j} X_{ij} P_R + m_{u_i} Y_{ij}P_L) d_j H^+ \frac{\sqrt{2}}{v} m_{\ell_j} Z_{ij} \bar{\nu}_L  \ell_R  H^+  + h.c.  \bigg),
\end{eqnarray}
where $X_ij$, $Y_ij$ and $Z_{ij}$ are defined in \cite{Hernandez-Sanchez:2020vax}, all being  functions of the parameters $\chi's$,    in such a way that we can recover the usuals models with discrete symmetry when the latter are absent. In general, the fermion-fermion-Higgs couplings ($\phi f f$) in our model can be written as $g_{2HDM-III}^ {\phi ff}= g_{\rm 2HDM-with-discrete-symetry} ^{\phi ff} + \Delta g$, being $\Delta g$ the contribution of the Yukawa texture.  

\section{Numerical Analysis}

In this section, we present a brief report about some of the best Benchmark Points (BPs) available over the 2HDM-III parameter space in its lepton-specific incarnation. Full details about the numerical analysis can be found in \cite{Hernandez-Sanchez:2020vax}.
\begin{figure}[h]
\begin{center}
\includegraphics[scale=0.7]{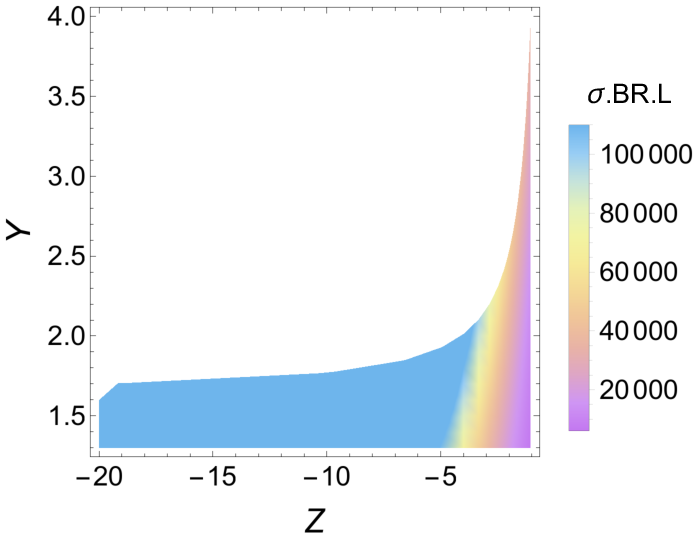}
\end{center}
\caption{Event rates for $\sigma (b\bar c\to H^+)\times BR(H^+\tau \nu_{\tau})  \times L$ (with $L=36.1\ fb^{-1}$) considering $\cos(\beta-\alpha)=0.5$,  $\{\chi_{22}^u,\chi_{23}^u,\chi_{33}^u,\chi_{22}^d,\chi_{23}^d,\chi_{33}^d,\chi_{22}^l,\chi_{23}^l,\chi_{33}^l\}=\{1,0.1,1.4,1.8,0.1,1.2,-0.4,0.1\}$,  $M_h=125$ GeV, $M_A=100$ GeV, $M_H=150$ GeV and $M_{H^+}=120$ GeV, in the $Z-Y$ plane for 2HDM-III like X. We assume $\sqrt s= 13$ TeV and $L=36.1$ fb$^{-1}$.  (From \cite{Hernandez-Sanchez:2020vax}.)  }
\label{events}
\end{figure}
In figure~\ref{events} we estimate the number of events $b\bar c\to H^+\to \tau \nu_{\tau}$ with respect to the $Z$ and $Y$ parameters.  The coloured regions represent parameter space that satisfies both theoretical and experimental constraints (for details, see \cite{Hernandez-Sanchez:2020vax}), specifically assuming $M_{H^\pm}=120$ GeV.  From these, we select one BP with $Y=1.6$ and $Z=-20$ ($X=-1/Z=0.04$),  which produces around two million events at the LHC for current energy and luminosity.  The main irreducible background comes from  $q\bar q\to W^{\pm} \to \tau \nu_{\tau}$.  However, the reducible backgrounds  $gp\to W^{\pm}q$ and $q\bar q\to W^+W^-$ have to be considered too. 

\begin{figure}
\begin{center}
\includegraphics[scale=0.7]{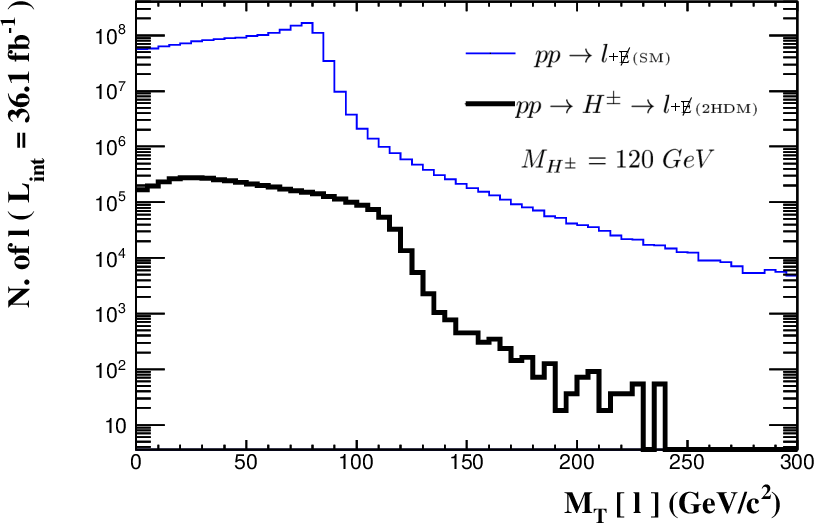}
\caption{Transverse mass for signal and  background without detector level cuts.  (From \cite{Hernandez-Sanchez:2020vax}.) }
\label{mass1}
\end{center}
\end{figure}

We then proceed to a full MC analysis, by using CalcHEP 3.7 \cite{Belyaev:2012qa} as an events generator,  PYTHIA6 \cite{Sjostrand:2006za} for parton shower and PGS \cite{PGS} as detector emulator (all details can be found \cite{Hernandez-Sanchez:2020vax}). Further, the final state particle kinematics was mapped with the help of MadAnalysis5 \cite{Conte:2012fm}.  Because we look for leptonic $\tau$ decays, multiple neutrinos in the final state prevent us from reconstructing its invariant mass.  Thus, we need to rely on the transverse mass to access the $H^\pm$ mass, so that  figure \ref{mass1} presents its shape for both signal and (total) background.  In this plot, is possible to see a Jacobian peak associated with the charged Higgs boson mass, to which it can be fit.   However, the background produced a much large number of events.  Therefore, it is necessary to impose a set of cuts to extract our signal.

\begin{figure}[h]
\begin{center}
\includegraphics[scale=0.7]{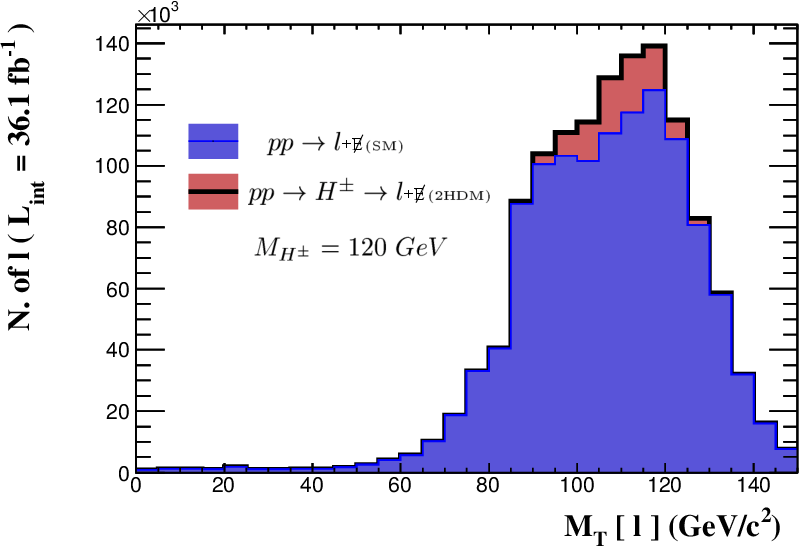}
\end{center}
\caption{Transverse mass for signal and  background after impose the cuts.   (From \cite{Hernandez-Sanchez:2020vax}.) }
\label{MASS2}
\end{figure}

For the purpose of optimising our Signal ($S$) versus background ($B$) ratio, the first important condition to exploit is a jet veto in the final state. This initial cut reduces the background by half, while more than $60\%$ of the signal is preserved.  For the rest of the events, we demand the following: $p_T(l)>45$ GeV,  $40$ GeV$< E\hspace{-2.3mm} / _T <70$ GeV, $|\eta(l)|<1.2$ and  $E_T>55$ GeV.  After imposing this set of cuts, the behaviour of the transverse mass is shown in figure \ref{MASS2}. The final selection is over such a transverse mass.  For the targeted $H^\pm$ mass,  we demand $85$ GeV$<M_T(l)<125$ GeV.  

\begin{table}
\begin{center}
\footnotesize
\begin{tabular}{|c|c|c|c|c|c|}
\hline 
 & Cut 1: $p_T(k)$ & Cut 2: $ E\hspace{-1.9mm} / _T$ & Cut 3: $|\eta(l)|$&Cut 4: $E_T$& Cut 5: $M_T(l)$ \\ 
\hline
\hline 
Signal  & 294136 & 237167 & 215684 &85480 & 82147 \\ 
\hline
Background  & 27527568 & 7919086 & 3832807 &1090294 &795470  \\ 
\hline
\end{tabular}
\caption{Number of events after imposing the cuts described in the text.  The significance is ${\cal{S}}=S/\sqrt{S+B}=87.68$. (From \cite{Hernandez-Sanchez:2020vax}.) }
\end{center}
 \label{numbers}
\end{table}
The numerical results after imposing  the described cuts are found in table 1.

\section{Conclusions}

To conclude, we believe it possible to extract a charged Higgs boson signal at the LHC within the 2HDM-III scenario in its lepton-specific incarnation, by searching for  $b \bar{c} \to H^-  \to \tau \bar \nu_\tau$, with the $\tau$  identified via decays into electrons/muons and corresponding neutrinos. If not already with present data for light charged Higgs bosons (as exemplified here), this could well happen  by the end of Run 3 over a $H^\pm$ mass  interval ranging from 100 GeV or so up to the TeV scale, as seen in Ref.~\cite{Hernandez-Sanchez:2020vax}. However, in order to achieve this, a dedicated selection procedure is required to be optimised around a tentative charged Higgs boson mass value, in order to suppress backgrounds effectively, both reducible and irreducible ones. We have obtained such results through a sophisticated MC analysis down to the detector level, so that we believe these to be solid enough to deserve further investigations by ATLAS and CMS.


\begin{thebibliography}{99}


\bibitem{Branco:2011iw} 
  G.~C.~Branco, P.~M.~Ferreira, L.~Lavoura, M.~N.~Rebelo, M.~Sher and J.~P.~Silva,
  Phys.\ Rept.\  {\bf 516}, 1 (2012)
  [arXiv:1106.0034 [hep-ph]].
  
\bibitem{Fritzsch:2002ga} 
  H.~Fritzsch and Z.~z.~Xing,
  Phys.\ Lett.\ B {\bf 555}, 63 (2003)
  [hep-ph/0212195].

\bibitem{DiazCruz:2009ek} 
  J.~L.~D\'{\i}az-Cruz, J.~Hern\'andez--S\'anchez, S.~Moretti, R.~Noriega-Papaqui and A.~Rosado,
  Phys.\ Rev.\ D {\bf 79}, 095025 (2009)
  [arXiv:0902.4490 [hep-ph]].

\bibitem{Hernandez-Sanchez:2016vys} 
  J.~Hern\'andez-S\'anchez, O.~Flores-S\'anchez, C.~G.~Honorato, S.~Moretti and S.~Rosado,
  PoS CHARGED {\bf 2016}, 032 (2017)
  [arXiv:1612.06316 [hep-ph]].

\bibitem{Hernandez-Sanchez:2015bda} 
  J.~Hern\'andez-S\'anchez, S.~P.~Das, S.~Moretti, A.~Rosado and R.~Xoxocotzi-Aguilar,
  PoS DIS {\bf 2015}, 227 (2015)
  [arXiv:1509.05491 [hep-ph]].

\bibitem{Das:2015kea} 
  S.~P.~Das, J.~Hern\'andez-S\'anchez, S.~Moretti, A.~Rosado and R.~Xoxocotzi,
  Phys.\ Rev.\ D {\bf 94}, no. 5, 055003 (2016)
  [arXiv:1503.01464 [hep-ph]].

\bibitem{Cordero-Cid:2013sxa} 
  A.~Cordero-Cid, J.~Hern\'andez-S\'anchez, C.~G.~Honorato, S.~Moretti, M.~A.~P\'erez and A.~Rosado,
  JHEP {\bf 1407}, 057 (2014)
  [arXiv:1312.5614 [hep-ph]].

\bibitem{Felix-Beltran:2013tra} 
  O.~F\'elix-Beltr\'an, F.~Gonz\'alez-Canales, J.~Hern\'andez-S\'anchez, S.~Moretti, R.~Noriega-Papaqui and A.~Rosado,
  Phys.\ Lett.\ B {\bf 742}, 347 (2015)
  [arXiv:1311.5210 [hep-ph]].

\bibitem{HernandezSanchez:2013xj} 
  J.~Hern\'andez-S\'anchez, S.~Moretti, R.~Noriega-Papaqui and A.~Rosado,
  PoS CHARGED {\bf 2012}, 029 (2012)
  [arXiv:1302.0083 [hep-ph]].
 
\bibitem{HernandezSanchez:2012eg} 
  J.~Hern\'andez-S\'anchez, S.~Moretti, R.~Noriega-Papaqui and A.~Rosado,
  JHEP {\bf 1307}, 044 (2013)
  [arXiv:1212.6818 [hep-ph]].

\bibitem{Akeroyd:2016ymd} 
  A.~G.~Akeroyd {\it et al.},
  Eur.\ Phys.\ J.\ C {\bf 77}, no. 5, 276 (2017)
  [arXiv:1607.01320 [hep-ph]].

\bibitem{Akeroyd:2012yg} 
  A.~G.~Akeroyd, S.~Moretti and J.~Hern\'andez-S\'anchez,
  Phys.\ Rev.\ D {\bf 85}, 115002 (2012)
  [arXiv:1203.5769 [hep-ph]].
  
\bibitem{Slabospitsky:2002gw}
    Slabospitsky, S. R.,
    CMS-NOTE-2002-010
    (2002)
  [arXiv:0203094[hep-ph]] .

  
\bibitem{Hernandez-Sanchez:2020vax}
J.~Hern\'andez-S\'anchez, C.~G.~Honorato, S.~Moretti and S.~Rosado-Navarro,
Phys. Rev. D \textbf{102}, no.5, 055008 (2020)
doi:10.1103/PhysRevD.102.055008
[arXiv:2003.06263 [hep-ph]].
  
 
  
 
 
  
  
 


  

  
      
  
  
   
  
 
 
  
  
  
  
  
 



  
\bibitem{Belyaev:2012qa} 
  A.~Belyaev, N.~D.~Christensen and A.~Pukhov,
  Comput.\ Phys.\ Commun.\  {\bf 184}, 1729 (2013)
  [arXiv:1207.6082 [hep-ph]].




\bibitem{Sjostrand:2006za} 
  T.~Sj{\"o}strand, S.~Mrenna and P.~Z.~Skands,
  JHEP {\bf 0605}, 026 (2006)
  [hep-ph/0603175].

\bibitem{PGS}
J. Conway, R. Culbertson, R. Demina, B. Kilminster, M. Kruse, S. Mrenna, J. Nielsen, M. Roco, A. Pierce, J. Thaler and T. Wizansky,\\
http://conway.physics.ucdavis.edu/research/software/pgs/pgs4-general.htm.


\bibitem{Conte:2012fm} 
  E.~Conte, B.~Fuks and G.~Serret,
  Comput.\ Phys.\ Commun.\  {\bf 184}, 222 (2013)
  [arXiv:1206.1599 [hep-ph]].
\end{thebibliography}
\end{document}